\documentstyle[prd,aps]{revtex}

\begin{document}
\draft

\twocolumn[\hsize\textwidth\columnwidth\hsize\csname
@twocolumnfalse\endcsname
\renewcommand{\theequation}{\thesection . \arabic{equation} }
\title{\bf Nonorientable spacetime tunneling}

\author{ Pedro F. Gonz\'{a}lez-D\'{\i}az and Luis J. Garay}
\address{Centro de F\'{\i}sica ``Miguel Catal\'{a}n'',
Instituto de Matem\'{a}ticas y F\'{\i}sica Fundamental,\\
Consejo Superior de Investigaciones Cient\'{\i}ficas,
Serrano 121, 28006 Madrid (SPAIN)}
\date{July 7, 1998}

\maketitle

\begin{abstract}

Misner space is generalized to have the nonorientable topology
of a Klein bottle, and it is shown that in a classical spacetime
with multiply connected space slices having such a topology,
closed timelike curves are formed. Different regions on the Klein
bottle surface can be distinguished which are separated by apparent
horizons fixed at particular values of the two angular variables
that enter the metric. Around the throat of this tunnel (which
we denote a Klein bottlehole), the position of these horizons
dictates an ordinary and exotic matter distribution such that,
in addition to the known diverging lensing action of wormholes,
a converging lensing action is also present at the mouths.
Associated with this matter distribution, the accelerating
version of this Klein bottlehole shows four distinct chronology
horizons, each with its own nonchronal region. A calculation
of the quantum vacuum fluctuations performed by using the
regularized two-point Hadamard function shows that each
chronology horizon nests a set of polarized hypersurfaces
where the renormalized momentum-energy tensor diverges. This
quantum instability can be prevented if we take the accelerating
Klein bottlehole to be a generalization of a modified Misner
space in which the period of the closed spatial direction is
time-dependent. In this case, the nonchronal regions and closed
timelike curves cannot exceed a minimum size of the order the
Planck scale.

\end{abstract}


\vskip2pc]

\renewcommand{\theequation}{\arabic{section}.\arabic{equation}}

\section{\bf Introduction}

Most of the hitherto proposed models for spacetime tunnelings
and time machines can be regarded as generalizations from
Misner space [1], obtained by replacing the planes of this
space with different orientable topologies, such as a sphere
in wormholes [2] and a torus in ringholes [3], or by
displacing by a suitable amount the period of the closed
spatial direction that distinguishes Misner from Minkowskii
space, as in Gott-Grant time machines [4,5]. Changing the
topology of Misner space preserves time-dilation as the
origin for the emergence of closed timelike curves (CTCs)
at suffiently late times in the nonchronal region, and offers
the possibility of obtaining exotic matter distributions
near the hole throat that allowed itineraries through the
tunnels along which an observer never finds any region
with negative energy density [3]; hence the observer could
travel safely from one mounth to the other.

It appears then of interest to investigate new kinds of
tunnelings with even more complicated topology. In
particular, nonorientable topologies would be specially
suited as they might give rise to sufficiently large
interior regions filled with ordinary matter only. In
this paper we will construct a spacetime tunnel with
the topology of a Klein bottle, and discuss the
properties of the possible time machines that can be
built out of it.

The interest of this research would be further increased
if we take into account the recent developments independently
advanced by Li and Gott [6] and by Gonz\'{a}lez-D\'{\i}az [7],
 according
to which Misner space becomes stable to quantum fluctuations
even on the chronology (Cauchy) horizon for a convenient
re-definition of its periodicity properties, and hence of its
associated vacuum. This would be a violation of chronology
protection [8] which should occur in all alluded generalizations
from Misner space and lended physical support to the recently
considered observable effects that spacetime tunnelings would
produce if they naturally existed in some sufficiently early
regions of the universe [9-11]. Clearly, more complicated
topologies such as that of the Klein bottle would quantitative
modify the observable predictions from cosmic wormholes and
ringholes [10] and are likely to induce new observable effects.
Moreover, the proposed study could also lend extra interest to
the proposal that the universe was not created from nothing, but
it created itseft by means of primordial CTCs [12]. However,
whereas the work by Li and Gott [6] would imply the possibility
for the existence of time machines and CTC's with macroscopic
sizes and large travels for which the concept of chronology
horizon is lost in the semiclassical treatment [13], the other
stabilization procedure [7] implies a well-behaved quantization
of time and chronology horizon themselves that only allows the
existence of time machines and CTC's with essentially Planck
size (see also Refs. [14,15] for a more balanced discussion).

Thus, the main motivation for the study of nonorientable
spacetime tunnelings is related to the notion of the so called
quantum time machine [7,16] which, together with virtual black
holes [17] and Euclidean wormholes [18], appears as a necessary
ingredient for a consistent description of the quantum spacetime
foam [19]. It seems a natural requirement that the foam should
contain all possible topologies, including nonorientable ones.
Actually, nonorientability may become a topological necessity if
occurrence of a nonzero minimum time and length at the Planck
scale are taken to be the hallmark of quantum spacetime foam
[20]. Whereas the former limit would ultimately imply existence
of causality-violating quantum time machines in the foam [19],
the latter one would lead to the topological impossibility of
keeping two-sideness for any two-surface lying in ${\bf R}^{3}$.
Since one of the two possible sides of such surfaces can always
be made topologically inaccessible by the uncertainty
$\bigtriangleup x\geq$ Planck length in the foam, any closed
two-surfaces (that is, any two manifold) lying in the foam
should necessarily be one-sided and hence nonorientable.

The paper is organized as follows: Using a given geometric
ansatz, in Sec. II we describe a way to obtain the static metric
on some sections of the spacetime generated by a distribution of
matter with the topology of a Klein bottle, and discuss the
existence of apparent horizons at fixed values of the angular
variables that define a nonorientably deformed toroidal
geometry. Starting with the metrics obtained for the Klein
bottle sections, we derive in Sec. III a spacetime that
describes what we may call a Klein bottlehole, that is a tunnel
in Lorentz spacetime with the symmetry of a Klein bottle,
discussing the conditions required by this tunnel to be embedded
in flat space, so as the characteristics of the stress-energy
tensor needed to make it compatible with general relativity and
the lensing actions expected to be induced in its mouths. Sec.
IV contains a discussion of the conversion of this Klein
bottlehole into time machine, i.e., into an accelerating Klein
bottlehole, briefly analysing the causal and noncausal structure
of the resulting space. Also in Sec. IV is a calculation of the
quantum effects implied by vacuum polarization inside the
chronology horizons, following the procedure used by Kim and
Thorne [21], so as a brief discussion of the above spacetime
construct in the case in which the period of the closed spatial
direction is time-dependent, for which case no polarized
hypersurfaces with divergent vacuum polarization are allowed.
Finally, we summarize our results in Sec. V.

\section{\bf Static metric on the Klein bottle}
\setcounter{equation}{0}

We shall consider the gravitational field created by a
distribution of matter with the symmetry of a Klein bottle,
and obtain the static spacetime metric for constant
surfaces possessing such a symmetry. In order to account
for the nonorientable character of the Klein bottle, we
shall extend the range of the angular coordinate $\varphi_1$
$0\leq\varphi_1\leq 2\pi$ on the circular axis of the
orientable torus [3] to also encompass the values continuously
running from $2\pi$ to $3\pi$, while allowing the radii of
the transversal section of the so-deformed torus tube and of
its deformed axis to be both $\varphi_1$-dependent,
with the transversal surfaces at
$\varphi_1=3\pi$ and at $\varphi_1=0$ identified.
In Fig. 1 we define
the Cartesian coordinates on a Klein bottle.
These can be written as
$x=m_1\sin\varphi_1$, $y=m_1\cos\varphi_1$ and $z=b_1\sin\varphi_2$
for $0\leq\varphi_1\leq 2\pi$, and
$x=m_2\sin\varphi_1$, $y=A_1-C_2 -m_2\cos\varphi_1$ and
$z=b_2\sin\varphi_2$, for $2\pi\leq\varphi_1\leq 3\pi$, with
\begin{equation}
m_1=a_1-b_1\cos\varphi_2 ,\;\; n_1=b_1-a_1\cos\varphi_2 ,
\end{equation}
where $0\leq\varphi_2\leq 2\pi$, and we have used the ansatz:
\begin{equation}
a_1\equiv a_1(\varphi_1)=(A_1-C_1)\cos^2\frac{\varphi_1}{4}+C_1
\end{equation}
\begin{equation}
b_1\equiv b_1(\varphi_1)=(B_1-D_1)\cos^2\frac{\varphi_1}{4}+D_1 ,
\end{equation}
in which $A_1,B_1,C_1$ and $D_1$ are adjustable constant parameters
satisfying the conditions $A_1 >C_1$, $B_1 >D_1$, $A_1 >B_1$,
and $C_1 >D_1$, with $a_1$ and $b_1$ the varying radius of
the circumference generated by the Klein bottle axis and
that of the transversal section of the Klein bottle tube,
respectively, in the region $0\leq\varphi_1\leq 2\pi$. For
the interval $2\pi\leq\varphi_1\leq 3\pi$, we have
\begin{equation}
m_2=a_2+b_2\cos\varphi_2 ,\;\; n_2=b_2+a_2\cos\varphi_2 ,
\end{equation}
for the associated ansatz
\begin{equation}
a_2\equiv a_2(\varphi_1)=(C_2-A_2)\sin^2\frac{\varphi_1}{2}+A_2
\end{equation}
\begin{equation}
b_2\equiv b_2(\varphi_1)=(D_2-B_2)\sin^2\frac{\varphi_1}{2}+B_2 ,
\end{equation}
where the conditions for the new adjustable constant parameters
are: $C_2 >A_2$, $D_2 >B_2$, $C_2 >D_2$,
and $A_2 >B_2$, and $D_2=B_1$, $B_2=D_1$, $A_1-C_1=A_2+C_2$,
with $A_1-C_1 >2A_2$.

We have in the region $0\leq\varphi_1\leq 2\pi$:
\[d\Omega_1^2=dx^2+dy^2+dz^2=\]
\[\left\{m_1^2+\frac{1}{4}\left[M_1(a_1-C_1)+N_1(b_1-D_1)\right]\right\}d\varphi_1^2\]
\begin{equation}
+b_1^2d\varphi_2^2-b_1\sqrt{(a_1-C_1)(A_1-a_1)}\sin\varphi_2 d\varphi_1 d\varphi_2 ,
\end{equation}
in which
\begin{equation}
M_1=A_1-a_1-(B_1-b_1)\cos\varphi_2
\end{equation}
\begin{equation}
N_1=B_1-b_1-(A_1-a_1)\cos\varphi_2 ,
\end{equation}
and, in the region $2\pi\leq\varphi_1\leq 3\pi$,
\[d\Omega_2^2=dx^2+dy^2+dz^2=\]
\[\left\{m_2^2+M_2(a_2-A_2)+N_2(b_2-B_2)\right\}d\varphi_1^2\]
\begin{equation}
+b_2^2d\varphi_2^2-2b_2\sqrt{(a_2-A_2)(C_2-a_2)}\sin\varphi_2 d\varphi_1 d\varphi_2 ,
\end{equation}
where
\begin{equation}
M_2=C_2-a_2+(D_2-b_2)\cos\varphi_2
\end{equation}
\begin{equation}
N_2=D_2-b_2+(C_2-a_2)\cos\varphi_2 .
\end{equation}

We can assume then for the static metric corresponding to a
distribution of matter with the symmetry of a Klein bottle
the general expression:
\[ds^2=-e^{\Phi}dt^2+e^{\Psi}dr^2\]
\begin{equation}
+\theta(2\pi-\varphi_1)d\Omega_1^2
+\theta(\varphi_1-2\pi)d\Omega_2^2 ,
\end{equation}
in which the $\theta(x)$'s are the step Heaviside function, with
$\theta(x)=1$ for $x>0$ and $\theta(x)=0$ for $x<0$,
\begin{equation}
r=r_1=\sqrt{a_1^2+b_1^2-2a_1 b_1\cos\varphi_2} ,
\end{equation}
for $0\leq\varphi_1\leq 2\pi$,
\begin{equation}
r=r_2=\sqrt{a_2^2+b_2^2+2a_2 b_2\cos\varphi_2} ,
\end{equation}
for $2\pi\leq\varphi_1\leq 3\pi$, and $\Phi$ and $\Psi$ will
generally depend on $t$ and $r$ in the respective interval
of $\varphi_1$.

Denoting by $x^0$, $x^1$, $x^2$ and $x^3$, respectively, the
coordinates on the Klein bottle $ct$, $r$, $\varphi_1$ and
$\varphi_2$, we have for the nonzero components of the metric
tensor: $g_{00}=-e^{\Phi}$, $g_{11}=e^{\Psi}$,
\[g_{22}=\]
\[\left\{m_1^2+
\frac{1}{4}\left[M_1(a_1-C_1)+N_1(b_1-D_1)\right]\right\}
\theta(2\pi-\varphi_1)\]
\[+\left[m_2^2+M_2(a_2-A_2)+N_2(b_2-B_2)\right]\theta(\varphi_1-2\pi)\]
\[g_{23}=
-\left(b_1\sqrt{(a_1-C_1)(A_1-a_1)}\theta(2\pi-\varphi_1)\right.\]
\[\left.+2b_2\sqrt{(a_2-A_2)(C_2-a_2)}\theta(\varphi_1-2\pi)\right)\sin\varphi_2\]
\[g_{33}=b_1^2\theta(2\pi-\varphi_1)+b_2^2\theta(\varphi_1-2\pi) .\]
Using then the expressions for the derivatives that result from
our previous definitions and ans"tze, i.e.:
\begin{equation}
\frac{dm_i}{dr_i}=\frac{r_i}{a_i}, \;\;\;
\frac{dn_i}{dr_i}=\frac{r_i}{b_i},\;\;\;
\frac{n_i}{m_i}=\frac{a_i}{b_i}
\end{equation}
\begin{equation}
\frac{dm_i}{d\varphi_2}=-(-1)^{i}b_i\sin\varphi_2 ,\;\;\;
\frac{dn_i}{d\varphi_2}=-(-1)^{i}a_i\sin\varphi_2
\end{equation}
\begin{equation}
\frac{da_i}{dr_i}=\frac{r_i}{m_i} ,\;\;
\frac{db_i}{dr_i}=\frac{r_i}{n_i}
\end{equation}
\begin{equation}
\frac{da_i}{d\varphi_1}=
\frac{i}{2}(-1)^{i}\sqrt{(a_i-C_i)(A_i-a_i)} ,\;\;
\end{equation}
\begin{equation}
\frac{db_i}{d\varphi_1}=
\frac{i}{2}(-1)^{i}\sqrt{(b_i-D_i)(B_i-b_i)} ,
\end{equation}
where $i=1,2$, one can calculate the components of the affine
connection and hence those of the Ricci curvature tensor.
From the resulting Einstein equations that correspond to
the gravitational field of a matter distribution with the symmetry
of the Klein bottle used in this paper, one finally obtains:
\[\frac{8\pi G}{c^4}\left(T_0^0-T_1^1\right)e^{\Psi}\]
\begin{equation}
=-\frac{1}{4}(\Psi '+\Phi ')P+\frac{1}{2}P'+\frac{1}{4}P^2+\frac{1}{2}Q
\end{equation}
\[\frac{8\pi G}{c^4}\left(T_2^2-T_3^3\right)e^{\Psi}\]
\begin{equation}
=\frac{1}{4}(\Psi '-\Phi ')R
-\frac{1}{2}R'+\frac{1}{2}(\ln g_{23})'R
-\frac{1}{2}RS+Te^{\Psi} ,
\end{equation}
where the prime denotes derivative with respect to the corresponding
coordinate $r_i$, and
\begin{equation}
P=(\ln g_{22})'+2(\ln g_{23})'+(\ln g_{33})'
\end{equation}
\begin{equation}
Q=\left[g^{22}g^{33}+(g^{23})^2\right]
\left[g_{22}'g_{33}'-(g_{23}')^2\right]
\end{equation}
\begin{equation}
R=(\ln g_{22})'-(\ln g_{33})' ,\;\;
S=(\ln g_{22})'+(\ln g_{33})'
\end{equation}
\[T=\frac{1}{2}g^{33}\left(\frac{dg^{22}}{d\varphi_2}\frac{dg_{23}}{d\varphi_1}
-\frac{dg^{22}}{d\varphi_1}\frac{dg_{23}}{d\varphi_2}
+2\frac{dg^{23}}{d\varphi_2}\frac{dg_{33}}{d\varphi_1}\right)\]
\[-g^{22}\left(\frac{1}{2}\frac{dg^{33}}{d\varphi_1}\frac{dg_{23}}{d\varphi_2}
+\frac{dg^{23}}{d\varphi_1}\frac{dg_{22}}{d\varphi_2}\right)\]
\[-\frac{1}{2}g^{22}(g^{23})^2\left[\left(\frac{dg_{22}}{d\varphi_2}\right)^2
+\frac{dg_{22}}{d\varphi_1}\frac{dg_{33}}{d\varphi_1}\right]\]
\begin{equation}
+\frac{1}{2}g^{33}(g^{23})^2\left(\frac{dg_{33}}{d\varphi_1}\right)^2 ,
\end{equation}
with the expressions for the metric tensor components as given above.

General solutions to these equations look quite complicated,
even for the vacuum case, $T_{\mu}^{\nu}=0$. Nevertheless,
one can still derive solutions to (2.21) and (2.22) on
two-dimensional sections in the vacuum case. Thus, on the
sections $\varphi_1=$Const. or on the sections $\varphi_2=$Const.
we can obtain solutions in closed form and investigate the
possible horizons which can appear along the angular
coordinates. Let us first consider constant-$\varphi_1$
sections of the Klein bottle. In this case, we have
the solution
\begin{equation}
\Phi=0 ,\;\; \Psi=2\ln\left[\frac{4b_i r_i}{r_i^2-(a_i^2-b_i^2)}\right] ,
\;\; i=1,2  ,
\end{equation}
where we have chosen the integration constant to be zero
and $a_i$ and $b_i$ are constant.

Solution (2.27) is defined for $0\leq t\leq\infty$,
$0\leq\varphi_2\leq 2\pi$, $a_i-b_i\leq r_i\leq a_i+b_i$,
on constant $\varphi_1$-sections which are fixed either on
$0\leq\varphi_1\leq 2\pi$ when $i=1$, or on
$2\pi\leq\varphi_1\leq 3\pi$ when $i=2$, and describes the
spacetime geometry on the corresponding $\varphi_1$-section
of a Klein bottle with constant $a_i$ and $b_i$ radii,
generated by varying angle $\varphi_2$ only. The variation
of the metric with angle $\varphi_2$ is of interest in order
to determine the position of angular horizons. On $\varphi_2=\pi$,
$r_i=r_{i max}=a_i+b_i$, the metric becomes singular. As
$\varphi_2$ decreases from $\pi$ to $\varphi_2=\varphi_2^c=
\arccos\frac{b_i}{a_i}$,
$r_{i max}$ does to $\sqrt{a_i^2-b_i^2}$,
where the metric is singular again. It is also singular on
$\varphi_2=0$, where $r_i=r_{i min}=a_i-b_i$, and on
$\varphi_2=2\pi-\varphi_2^c$. All of these singularities
are not true singularities, but arise only from the choice
of coordinates. They are also present in the static metric
on a two-torus, which is the orientable topology that
directly corresponds to that of a Klein bottle.

If we introduce the new coordinates
\begin{equation}
U_i=t+2b_i\ln\left[r_i^2-(a_i^2-b_i^2)\right]
\end{equation}
\begin{equation}
V_i=t-2b_i\ln\left[r_i^2-(a_i^2-b_i^2)\right] ,
\end{equation}
then the metric transforms into:
\begin{equation}
ds^2=dU_i dV_i+\theta(2\pi-\varphi_1)d\Omega_1^2 +
\theta(\varphi_1-2\pi)d\Omega_2^2 ,
\end{equation}
where
\begin{equation}
r_i^2=a_i^2-b_i^2+\exp\left(\frac{U_i-V_i}{4b_i}\right).
\end{equation}
Metric (2.30) is in fact regular on the above horizons.

Consider now constant $\varphi_2$-sections of the Klein bottle.
In this case, we get the closed form vacuum solution: $\Phi=0$ and
\begin{equation}
\Psi=
\ln\left\{\frac{\frac{2m_i}{a_i}+\frac{i^2}{4}\left(-4+
\frac{\alpha_i}{a_i}+\frac{\beta_i}{b_i}+\frac{m_i^{(0)}}{m_i}
+\frac{n_i^{(0)}}{n_i}\right)r_i^2}{m_i^2+
\frac{i^2}{4}\left[M_i(a_i-\alpha_i)+N_i(b_i-\beta_i)\right]}
\right\},
\end{equation}
where $i=1,2$, $\alpha_{1,2}=C_1,A_2$, $\beta_{1,2}=D_1,B_2$,
and $m_i^{(0)}$ and $n_i^{(0)}$ are constant
\begin{equation}
m_i^{(0)}=\delta_i+(-1)^i\gamma_i\cos\varphi_2 ,\;\;
n_i^{(0)}=\gamma_i+(-1)^i\delta_i\cos\varphi_2 ,
\end{equation}
with $\delta_{1,2}=A_1,C_2$ and $\gamma_{1,2}=B_1,D_2$.

We note that there is no singularity on the surfaces
$r_i=\sqrt{a_i^2-b_i^2}$ since both the numerator and the
denominator of the $g_{11}$-component of the metric
tensor corresponding to solution
(2.32) go to zero on such surfaces. On the region
$2\pi\geq\varphi_1\geq0$ singularities would appear at
values of angle $\varphi_1$ given by
\begin{equation}
\varphi_1=
4\arccos\chi_1 ,
\end{equation}
where
\[\chi_1=\]
\begin{equation}
\sqrt{\frac{8m_1^1\bigtriangleup m_1
+W_1+4\sqrt{W_1(m_1^1\bigtriangleup m_1+\frac{W_1}{16}+(m_1^1)^2)}}
{2[W_1-4(\bigtriangleup m_1)^2]}} ,
\end{equation}
in which
\begin{equation}
m_1^1=C_1-D_1\cos\varphi_2 ,\;\; \bigtriangleup m_1=m_1^{(0)}-m_1^1
\end{equation}
\begin{equation}
n_1^1=D_1-C_1\cos\varphi_2 ,\;\; \bigtriangleup n_1=n_1^{(0)}-n_1^1
\end{equation}
\begin{equation}
W_1=\bigtriangleup m_1(A_1-C_1)+\bigtriangleup n_1(B_1-D_1) .
\end{equation}
It can now be checked that a singularity can only appear on this
region whenever $B_1-D_1 > A_1-C_1$, i.e. provided the radius
$b_1$ of the Klein bottle decreases more rapidly than the
internal radius $a_1$ does as one approaches $\varphi_1=2\pi$.

As for region $2\pi\leq\varphi_1\leq 3\pi$, singularities may
be present whenever
\begin{equation}
\varphi_1=
4\arcsin\chi_2 ,
\end{equation}
where
\[\chi_2=\]
\begin{equation}
\sqrt{\frac{2m_2^1\bigtriangleup m_2
+W_2\pm 2\sqrt{W_2(m_2^1\bigtriangleup m_2+\frac{W_2}{4}+(m_2^1)^2)}}
{2[W_2-(\bigtriangleup m_2)^2]}} ,
\end{equation}
in which
\begin{equation}
m_2^1=A_2+B_2\cos\varphi_2 ,\;\; \bigtriangleup m_2=m_2^{(0)}-m_2^1
\end{equation}
\begin{equation}
n_2^1=B_2+A_2\cos\varphi_2 ,\;\; \bigtriangleup n_2=n_2^{(0)}-n_2^1
\end{equation}
\begin{equation}
W_2=\bigtriangleup m_2(C_2-A_2)+\bigtriangleup n_2(D_2-B_2) .
\end{equation}
On this region the existence of singularities does not impose any
constraints on the values of the parameters that define the
geometry of the Klein bottle, unless that $A_2$ must vanish
on the singularities. Then, singularities would appear for
any values of $C_2$, $B_2$ and $D_2$, provided $A_2=0$,
satisfying the conditions that follow the ansatz (2.5) and
(2.6) only on the extreme, critical values $\varphi_1=2\pi$
and $\varphi_1=3\pi$.

\section{\bf The spacetime of a Klein bottle hole}
\setcounter{equation}{0}

More complicated, but similar to as it happens in ringholes [3],
the creation of traversible nonorientable holes respecting
Einstein equations, so that classical general relativity be
valid everywhere, should be accompanied by the formation at late
times of CTC's in some nonchronal spacetime domains, and by
violation of the averaged weak energy condition [2,22] only on
some restricted, classically forbidden regions bounded by the
angular horizons dealt with in the previous Section. On the
other hand, one should also expect that violation of the energy
condition would not ultimately induce any divergences of either
the expectation value of a propagating scalar field squared or
the renormalized stress-energy tensor if, besides replacing the
planes of Misner space for Klein bottles, one takes the period
associated with the closed spatial dimension to be
time-dependent and given by $a=2\pi t^2$ [7]. The result would
be a quantically stable spacetime tunneling possessing CTC's
only at the Planck scale. If, moreover, we would take that
period to be $a=2\pi$ and re-define the vacuum consequently as
Li and Gott have recently done [6], the resulting nonchronal
region will, in principle, not possess instabilities anywhere,
too, and would give rise to CTC's which are not restricted in
size. In this case, however, the concept of a chronology horizon
can be argued to be lost [13].

A static nonorientable hole having the topology of a Klein
bottle would be traversible if a two-Klein bottle surrounding
one of its mouths where spacetime is nearly flat can be regarded
as an outer trapped surface to an observer looking through the
hole from the other mouth [2,23]. The static spacetime metric
for one such single, traversible Klein bottlehole may, in
principle, be written in the form
\[ds^2=
-dt^2+\theta(2\pi-\varphi_1)\left[\left(\frac{n_1(l_1)}{r_1(l_1)}\right)^2 dl_1^2
+d\Omega_1^2(l_1)\right]\]
\begin{equation}
+\theta(\varphi_1-2\pi)\left[\left(\frac{n_2(l_2)}{r_2(l_2)}\right)^2 dl_2^2
+d\Omega_2^2(l_2)\right] ,
\end{equation}
where $-\infty <t<+\infty$, $-\infty <l_i<+\infty$, the $d\Omega_i^2$'s
are as given by (2.7) and (2.10), with $b_i$ replaced for
$\sqrt{l_i^2+b_{01}^2}$ and $l_i$ the proper radial distance of each
transversal section of the Klein bottle on the respective
$\varphi_i$-interval for $i$; $b_{0i}$ is as given by (2.3)
and (2.6) for constant parameters ajusted to the radius
of the double throat of the Klein bottle that occurs at
$l_i=0$. Consequently,
\begin{equation}
m_i(l_i)=a_i+(-1)^i\sqrt{l_i^2+b_{0i}^2}\cos\varphi_2
\end{equation}
\begin{equation}
n_i(l_i)=\sqrt{l_i^2+b_{0i}^2}+(-1)^i a_i\cos\varphi_2
\end{equation}
\begin{equation}
r_i(l_i)=\sqrt{a_i^2+l_i^2+b_{0i}^2+2(-1)^i\sqrt{l_i^2
+b_{0i}^2}a_i\cos\varphi_2} .
\end{equation}
Metric (3.1) would give us a particularly simple example
of a traversible nonorientable hole which can be readily
generalized. Thus, one can convert (3.1) into the more
general static Klein bottlehole metric
\[ds^2=-e^{2\Phi}dt^2
+\theta(2\pi-\varphi_1)\left(\frac{dr_1^2}{1-\frac{K(b_1)}{b_1}}
+d\Omega_1^2\right)\]
\[+\theta(\varphi_1-2\pi)\left(\frac{dr_2^2}{1-\frac{K(b_2)}{b_2}}
+d\Omega_2^2\right)=\]
\[-e^{2\Phi}dt^2+\theta(2\pi-\varphi_1)\left((\frac{n_1}{r_1})^2dl_1^2
+d\Omega_1^2\right)\]
\begin{equation}
+\theta(\varphi_1-2\pi)\left((\frac{n_2}{r_2})^2dl_2^2
+d\Omega_2^2\right)
\end{equation}
if we let $\Phi=0$, $K(b_i)=b_{0i}^2/b_i$ and
$l_i=\pm\sqrt{b_i^2-b_{0i}^2}$, where the minus sign applies
on the left side of the throat and the plus sign applies on
the right side [2]. $\Phi$ will generally be given now as a
function of the mass of the nonorientable Klein bottlehole
and the geometric parameters that determine it.

The metric (3.5) can be regarded as a generalization to
Klein bottle symmetry from the static metric of a toroidal
ringhole, and hence from that of a spherical wormhole metric.
One can obtain the line element for a ringhole spacetime
from (3.5) by using the following set of parameters:
$A_1=C_1\neq 0$, $B_1=D_1\neq 0$,
$A_2=C_2=B_2=D_2=0$, and from the ringhole
metric we obtain the line element of a spherical wormhole
by using the transformations $a\rightarrow 0$,
$\varphi_2\rightarrow\theta+\frac{\pi}{2}$ [3]. Thus, we
obtain first for the ringhole metric [3]
\begin{equation}
ds^2=-e^{2\Phi}dt^2+\left(\frac{n}{r}\right)^2 dl^2
+m^2 d\varphi_1^2+b^2 d\varphi_2^2 ,
\end{equation}
and then for the wormhole metric [2]
\begin{equation}
ds^2=-e^{2\Phi}dt^2+dl^2+r^2(d\theta^2+\sin^2\theta d\phi^2).
\end{equation}

However, in order for the metric (3.5) to represent tunneling
through a traversible nonorientable Klein bottlehole, while
satisfying Einstein equations for a convenient stress-energy tensor,
it must be embeddible in a three-dimensional Euclidean space at
a fixed time $t$. What one should actually consider for such a
purpose is a three-geometry respecting the symmetry of the Klein
bottle and satisfying $a_i\geq b_i >l_i$, visualizing then the
given slice as removed from the spacetime of the nonorientable
Klein bottlehole and embedded in three-dimensional Euclidean
space. As it stands, metric (3.5) is not the metric required
for such an embedding, at least if we take for the embedding
space a space with cylindrical coordinates $z$, $r$, $\phi$
\begin{equation}
ds^2=dz^2+dr^2+r^2 d\phi^2 .
\end{equation}
Nevertheless, since $r_i$ and $\varphi_1$ are not independent
of each other, one can always convert (3.5) into a metrical
form which is embeddible in the cylindrical space (3.8). In
order to do that conversion, we first obtain the formula that
expresses the way in which $r_i$ varies with $\varphi_1$, i.e.
\[\frac{dr_i}{d\varphi_i}\equiv Q(i)=\]
\begin{equation}
-\frac{\left[m_i(A_i-C_i)+n_i(B_i-D_i)\right]
\sin\left(\frac{i\varphi_1}{2}\right)}{2ir_i} .
\end{equation}
Hence,
\[ds^2=-e^{2\Phi}dt^2+\]
\[\theta(2\pi-\varphi_1)\left(\frac{c(1)dr_1^2}{1-\frac{b_{01}^2}{b_1^2}}
+d(1)Q(1)dr_1 d\varphi_1+d\Omega_1^2\right)\]
\begin{equation}
+\theta(\varphi_1-2\pi)\left(\frac{c(2)dr_2^2}{1-\frac{b_{02}^2}{b_2^2}}
+d(2)Q(2)dr_2 d\varphi_1+d\Omega_2^2\right),
\end{equation}
with $c(i)+d(i)=1$.

Taking $dz=\frac{dz}{dr_1}dr_1+\frac{dz}{d\varphi_1}d\varphi_1$ for
$\varphi_1\leq 2\pi$, or
$dz=\frac{dz}{dr_2}dr_2+\frac{dz}{d\varphi_1}d\varphi_1$ for
$\varphi_1 > 2\pi$, one can obtain for any value of the
coordinate $\varphi_2$,
\begin{equation}
c(i)=1+2(1-\frac{b_{0i}^2}{b_i^2})
-2\sqrt{1-\frac{b_{0i}^2}{b_i^2}}.
\end{equation}
Therefore, the metric for the nonorientable Klein bottlehole
which is embeddible in flat space is described by (3.10),
with $c(i)$ given by (3.11) and $d(i)=1-c(i)$. Using these
coefficients, metric (3.8) will be the same as metric (3.10)
for constant values of $\varphi_2$ if we identify the
coordinates $r,\phi$ of the embedding space with either the
coordinates $r_1,\varphi_1$, for $\varphi_1\leq 2\pi$, or
the coordinates $r_2,\varphi_1$, for $\varphi_1 >2\pi$, and
if we require the function $z$ to satisfy
\begin{equation}
\frac{dz}{dr_i}=1+\left(1-\frac{b_{0i}^2}{b_i^2}\right)^{-1}
-2\left(1-\frac{b_{0i}^2}{b_i^2}\right)^{-\frac{1}{2}},
\end{equation}
for any value of $\varphi_1$, and
\begin{equation}
\frac{dz}{d\varphi_1}
=\frac{1}{2}\sqrt{(R(\varphi_2)_1-r_1)(r_1-\rho(\varphi_2)_1)} ,
\end{equation}
for $\varphi_1\leq 2\pi$, and
\begin{equation}
\frac{dz}{d\varphi_1}
=\sqrt{(R(\varphi_2)_2-r_2)(r_2-\rho(\varphi_2)_2)} ,
\end{equation}
for $\varphi_1 >2\pi$, where
\begin{equation}
R(\varphi_2)_1=A_1-B_1\cos\varphi_2 ,\;\;
\rho(\varphi_2)_1=C_1-D_1\cos\varphi_2
\end{equation}
and
\begin{equation}
R(\varphi_2)_2=C_2+D_2\cos\varphi_2 ,\;\;
\rho(\varphi_2)_2=A_2+B_2\cos\varphi_2 .
\end{equation}

From these expressions and the requirement that nonorientable
Klein bottleholes be connectible to asymptotically flat
spacetime, one can deduce how the embeddible surfaces would
flare at or around the hole throat. Thus, from (3.12) one
obtains
\[\frac{d^2 r}{dz^2}=\]
\begin{equation}
\frac{b_{0i}^2 r_i}{b_i^3 n_i}
\left(\frac{1}{\sqrt{1-\frac{b_{0i}^2}{b_i^2}}}-1\right)
\left(1+\frac{1}{1-\frac{b_{0i}^2}{b_i^2}}
-\frac{2}{\sqrt{1-\frac{b_{0i}^2}{b_i^2}}}\right)^{-\frac{7}{2}} ,
\end{equation}
which is positive for $2\pi-\varphi_2^c >\varphi_2 >\varphi_2^c
=\arctan(\frac{b_i}{a_i})$, and negative for
$-\varphi_2^c <\varphi_2 <\varphi_2^c$. Thus, exactly as it happens
in the case of toroidal ringholes [3], the embedding surface
flares outward for $\frac{d^2 r}{dz^2}>0$, and flares inward
for $\frac{d^2 r}{dz^2}<0$.

To investigate how the embedding surface flares at or around
the throat as the angle $\varphi_1$ is varied, we have to
distinguish two cases. The first one corresponds to condition
(3.13), from which we can get
\[\frac{d^2\varphi_1}{dz^2}=\]
\begin{equation}
\frac{(-2r_1+R(\varphi_2)_1+\rho(\varphi_2)_1)
(R(\varphi_2)_1-\rho(\varphi_2)_1)\sin\frac{\varphi_1}{2}}
{2\left[(R(\varphi_2)_1-r_1)(r_1-\rho(\varphi_2)_1)\right]^2} .
\end{equation}

Since $a_1 >b_1$ for $0\leq\varphi_1\leq2\pi$, the sign of
the r.h.s. of (3.18) will be fixed by the sign of the
quantity in the first brackets in its numerator. One obtains
that (3.18) vanishes for $\varphi_1=\varphi_1^2=\pi$ and
becomes negative for $\varphi_1 <\pi$, for which angular values
the embedding surface flares toward larger values of
the radius $b_1$, and negative for $\varphi_1 >\pi$, on
which region the embedding surface flares toward smaller
values of $b_1$.

The second case comes from condition (3.14). Here we get
\[\frac{d^2\varphi_1}{dz^2}=\]
\begin{equation}
-\frac{(-2r_2+R(\varphi_2)_2+\rho(\varphi_2)_2)
(R(\varphi_2)_2-\rho(\varphi_2)_2)\sin\varphi_1}
{4\left[(R(\varphi_2)_2-r_2)(r_2-\rho(\varphi_2)_2)\right]^2} .
\end{equation}
The critical value of $\varphi_1$ becomes then
$\varphi_1=\varphi_1^c=\frac{5\pi}{2}$. Again it is the
quantity in the first brackets in the numerator of the
r.h.s. of (3.19) which determines the sign of this equation.
For $\varphi_1 <\frac{5\pi}{2}$, that sign is negative so that
the embedding surface flares toward smaller values of $b_2$,
while it becomes positive for $\varphi_1 >\frac{5\pi}{2}$,
where the embedding surface flares toward larger values of $b_2$.

On the other hand, from the Einstein equations (2.21),
on the region $0\leq\varphi_1\leq2\pi$,
we can
obtain for the metric components of metric (3.10) with
$\Phi=0$,
\[\frac{\frac{8m_1}{a_1}-\left(\frac{A_1-C_1}{a_1}
+\frac{B_1-D_1}{b_1}\right)\cos^2\frac{\varphi_1}{4}+\frac{m_1^{(0)}}{m_1}+
\frac{n_1^{(0)}}{n_1}-2}{4\left\{n
m_1^2+\frac{1}{4}\left[M_1(A_1-C_1)
+N_1(B_1-D_1)\right]\cos^2\frac{\varphi_1}{4}\right\}}\]
\[+\frac{2(1+\sin\varphi_2)}{n_1 b_1}
-\frac{4\cos\frac{\varphi_1}{2}\sin\varphi_2}{(A_1-C_1)m_1
\sin^2\frac{\varphi_1}{2}}=Y_1(\varphi_1,\varphi_2)\]
\begin{equation}
=\frac{8\pi G}{c^4 r_1}(T_0^0-T_1^1) .
\end{equation}
The stress-energy tensor components $T_0^0$ and $T_1^1$ in
Eqn. (3.20) cannot be directly expressed in terms of,
respectively, an energy density $\rho$ and a tension per
unit area $\sigma$ for the symmetry associated with a
Klein bottle. Such as it happened for a toroidal symmetry [3],
here the tensor components $T_{\mu}^{\nu}$ will also depend
explicitly on $r_1$, whereas $\rho$ and $\sigma$ should be defined
as a function of the (nonorientable) normal to the surface element
on the Klein bottle, on the region where $0\leq\varphi_1\leq 2\pi$,
along the direction determined by the radius $b_1$. Since
$db_1/dr_1=r_1/n_1$, in the neighborhood of the throat where
$b_1\simeq b_{01}$, we must have
\begin{equation}
\rho c^2-\sigma=\left(\frac{n_1}{r_1}\right)^3(T_0^0-T_1^1)
\simeq\frac{c^4 b_{01}^2 n_1^2}{16\pi Gb_1^3 r_1^2}Y_1(\varphi_1,
\varphi_2) .
\end{equation}
Now, since the factor in front of $Y_1$ in (3.21) is positive
definite, it follows:
\begin{equation}
{\rm sgn}\left[\rho c^2-\sigma\right]
{\rm sgn}Y_1(\varphi_1,\varphi_2),
\end{equation}
at or near the nonorientable hole throat. An analysis of the
function $Y_1(\varphi_1,\varphi_2)$ indicates that $\rho c^2
-\sigma$ will be negative for small values of the involved
angles $\varphi_1$ and $\varphi_2$, and positive as one
approaches either $\varphi_1=2\pi$ or $\varphi_2=\pi$.
There will be then intermediate critical values for these angles
at which $Y_1=0$. These critical values will depend on the
values of the adjustable parameters that define the radii
$a_1$ and $b_1$. A similar analysis can be made for the
region $2\pi\leq\varphi_1\leq3\pi$, at or near the throat,
which allows us to conclude that the new function $Y_2$
will be positive for values of $\varphi_1$ close to $2\pi$,
and becomes negative as $\varphi_1$ approaches $3\pi$,
having the same behaviour as $Y_1$ with respect to
variation with angle $\varphi_2$. All of these results
have been obtained for the specific metric where $\Phi=0$,
but it is easy to check that they are still valid for
any other value of $\Phi$, provided it is everywhere finite.
It follows that for an observer moving through the Klein
bottlehole's throat with a sufficiently large speed,
$\gamma >>1$, the energy density
$\gamma^2(\rho c^2-\sigma)+\sigma$ will take on positive
or negative values depending on the specific combination
of values he chooses for $\varphi_1$, $\varphi_2$,
$A_i$, $B_i$, $C_i$ and $D_i$.

One would expect lensing effects to occur on the mouths of
the nonorientable Klein bottlehole with respect to a bundle
of light rays, at or near the throat, coming from the
distribution of positive/negative values for the energy
density; i.e.: the mouths would act like a diverging lens
for world lines along the values of the coordinates, at
or near the throat, which
correspond to negative energy density, and like a
converging lens for world lines passing through regions
with positive energy density. Thus, at or near the throat
of the Klein bottlehole, one would expect diverging lens
effects to tend to be concentrated onto those values of
$\varphi_1$ for which the radius of the transversal
section of the Klein bottle becomes larger, and on the
regions described by values of $\varphi_2$ which tend
to concentrate about $\varphi_2=\pi$. The exact relative
extend of such regions will ultimately depend on the
precise values used for the constant parameters that
define the radii $a_i$ and $b_i$. Actually, in order
to acertain with full accuracy which regions around the
throat behave like a lens a way or another, one should
consider the null-ray propagation governed by the integral
of the stress-energy tensor. For the mouths to defocus a
bundle of rays, such an integral,
\[\int_{l_i^1}^{\infty}dl_i e^{-\Phi}(\rho c^2-\sigma), \]
must turn out to be negative for any $l_i^1 <0$, and
positive if the mouths focus the rays. By using expressions
such as (3.21), one can check the above conclusions, for any
$\Phi$ which is everywhere finite.

\section{\bf Nonorientable time machine and vacuum fluctuations}
\setcounter{equation}{0}

The nonorientable Klein bottlehole considered in Sec. III
can be viewed as a generalization from Misner space, obtained by
replacing the identified flat planes of this space for
identified Klein bottles. It actually represents a static
tunneling between two asymptotically flat regions when we
give these Klein bottles vanishing relative velocity, $v=0$,
and is equivalent to extract two Klein bottles, with geometric
parameters given by (2.2), (2.3), (2.5) and (2.6), from
three-dimensional Euclidean space, and identify the Klein
bottle surfaces, so that when you enter the surface of, say
the right Klein bottle, you find yourself emerging from the
surface of the left Klein bottle, and vice versa. In Minkowski
spacetime, the Klein bottlehole can then be obtained identifying
the two world nonorientable concentric tube pairs swept out by
the two Klein bottles, with events at the same Lorentz time
identified.

Converting this Klein bottlehole into time machine is very
simple: one sets one of the nonorientable hole mouths in
motion at a given speed realtive to the other mouth, identifying
then the two Klein bottlehole's mouths to each other. We shall
consider now the spacetime metric of the resulting accelerating
Klein bottlehole. Let us assume the right mouth to be the mouth
which is moving. Then, just outside the right asymptotic rest
frame, the transformation of the Klein bottlehole coordinates
into external, Lorentz coordinates with metric
\[ds^2=-dT^2+\sum_{\alpha=1}^{3}dX_{i\alpha}^2 ,\;\; i=1,2 \]
can be given as
\begin{equation}
T_i=T_R+v\gamma l_i\sin\varphi_2,\;\;
X_{i3}=X_{3R}+\gamma l_i\sin\varphi_2
\end{equation}
\begin{equation}
X_{i1}=m_i(l_i)\sin\varphi_1 ,\;\;
X_{i2}=m_i(l_i)\cos\varphi_1 ,
\end{equation}
where $v=dX_{3R}/dT_R$ is the velocity of the right mouth;
$X_3=X_{3R}(t)$, $T=T_R(t)$ is the world line of the mouth's
center, $dt^2=dT_R^2-dX_{3R}^2$, and $\gamma$ is the
relativistic factor $\gamma=1/\sqrt{1-v^2}$. It follows that
just outside the left asymptotic rest frame, one should have
the transformation
\begin{equation}
T=t,\;\; X_{i3}=X_{3L}+l_i\sin\varphi_2 ,
\end{equation}
with the expressions for $X_{i1}$ and $X_{i2}$ also given by
(4.2). In (4.3), $X_{3L}$ is the time-independent $X_3$ location of
the left mouth's center of the Klein bottle. One can write then the
metric inside the accelerating Klein bottlehole  and outside but near
its mouths as:
\[ds^2=-e^{2\Phi}dt^2+\]
\[\theta(2\pi-\varphi_1)\left\{\left[-(1+gl_1 F(l_1)\sin\varphi_2)^2
+1\right]e^{2\Phi}dt^2\right.\]
\[\left.+c(1)dl_1^2+d(1)Q(1)dr_1 d\varphi_1+d\Omega_1^2
\right\}\]
\[+\theta(\varphi_1-2\pi)\left\{\left[-(1+gl_2 F(l_2)\sin\varphi_2)^2
+1\right]e^{2\Phi}dt^2\right.\]
\begin{equation}
\left.+c(2)dl_1^2+d(2)Q(2)dr_2 d\varphi_1+d\Omega_2^2
\right\}      ,
\end{equation}
where $g=\gamma^2 dv/dt$ is the acceleration of the right
mouth and $\Phi$ is the same function as for the original
static Klein bottlehole. The functions $F(l_i)$ are form
factors that vanish on the left half of the hole where
$l_i\leq 0$, rising monotonously from 0 to 1 as one moves
rightward from the throat to the right mouth [2]. Also used to
obtain metric (4.4) are the definitions: $dv=gdt/\gamma^2$,
$dt=dT_R/\gamma$ and $d\gamma=vg\gamma dt$.

Metric (4.4) is a specialization to nonorientable symmetry
from the metric used for accelerating toroidal ringholes [3].
Using for (4.4) the set of parameters $A_1=C_1\neq 0$,
$B_1=D_1\neq 0$, $A_2=C_2=B_2=D_2=0$, we in fact obtain the
metric for an accelerating ringhole [3]. Moreover, with the
additional
transformations $a\rightarrow 0$, $\varphi_2\rightarrow\theta
+\frac{\pi}{2}$, $\varphi_1\rightarrow\phi$, we finally get
the metric used by Morris {\it et al.} for accelerating
spherical wormholes [2], starting from (4.4).

At sufficiently late times, accelerating Klein bottleholes can
generate CTC's by exactly the same causes as in Misner or
accelerating wormhole and ringhole spaces [2,23]: on the left
mouth the Lorentz time and the proper time coincide, but on the
right mouth the latter time is relativistically dilated. When
this proper time shift becomes larger than the separation
between the hole mouths, then CTC's would appear. This happens
once the so-called chronology (Cauchy) horizon is reached. Such
a horizon is the onset of the nonchronal region and divides the
spacetime into two parts with completely different causal
properties. Like in Misner and accelerating wormhole and
ringhole spaces, there will be two families of timelike
geodesics in the chronal region of accelerating Klein bottlehole
space: rightward geodesics and leftward geodesics, both
possessing their own chronology horizons and nonchronal regions
[2,3]. All the mouth's lensing actions produced in accelerating
ringholes [3] are expected to occur in the present case as well,
including the drastic changes of the geometry of the chronology
horizon that originates, roughly speaking, a compact fountain
and a light cone at one of the hole's mouths [8,24]. Thus, if
you go through the Klein bottlehole along a given world line,
then one of the above chronology horizons and its nonchronal
region are destroyed. The chronology horizon is transformed into
just a boundary for the future Cauchy development of the compact
fountain, generated by null geodesics which are past directed,
to asymptote and enter the fountain [1]. All the effects caused
by this action are qualitatively similar to those caused in
accelerating ringholes [3] and, therefore, the reader interested
in more details on these effects is referred to Ref. [3].

Clearly, the acutest problem with the kind of time machines
being considered arises from the instabilities that such
spacetimes show when quantum vacuum polarization is taken into
account. In order to investigate what is going on, let us
consider the point-splitting regularized Hadamard two-point
function for a quantized, massless conformally coupled scalar
field propagating in the spacetime of an accelerating Klein
bottlehole. For regions where the curvature nearly vanishes
[25], the Hadamard function can now be written in the form
\[G_{reg,i}^{(1)\pm}(x,x')=
\sum_{N=1}^{\infty}\frac{\xi}{4\pi^2 D}
\left(\frac{b_i\xi}{2D}\right)^{N-1}\]
\begin{equation}
\times\left(\frac{1}{\lambda_{Ni}^{\pm}(x,x')}
+\frac{1}{\lambda_{Ni}^{\pm}(x',x)}\right) ,
\end{equation}
where
\begin{equation}
\xi=\sqrt{\frac{1-v}{1+v}} < 1 ,
\end{equation}
$D$ is the spatial length of a geodesic that connects points
$x$ and $x'$ by traversing once the Klein bottlehole, and
\begin{equation}
\lambda_{Ni}^{\pm}(x,x')=\xi^N\frac{\sigma_{Ni}^{\pm}}{\zeta_N} ,
\end{equation}
in which
\begin{equation}
\zeta_N=D\left(\frac{1-\xi^N}{1-\xi}\right)
\end{equation}
and $\sigma_{Ni}^{\pm}$ is the $N$th geodetic interval between $x$
and $x'$, for (+) $2\pi-\varphi_2^c >\varphi_2 >\varphi_2^c$ and
(-) $-\varphi_2^c <\varphi_2 <\varphi_2^c$, and ($i=1$)
$0\leq\varphi_1\leq 2\pi$, and $(i=2)$ $3\pi\geq\varphi_1\geq
2\pi$. $\lambda_{Ni}^{\pm}$ has been evaluated by means of a
generalization [21] of the method originally used by Hiscock and
Konkowski [25]. For the case of an accelerating Klein bottlehole
space, the use of Fig. 1 and the covering space which
distinguishes identified points in the original space [22]
allows us to compute the displacements at fixed times $T$ and
$T'$ of, respectively, copy 0 of $x'$ and copy $N$ of $x$ from
the covering-space throat location for $\varphi_1=0$ or
$\varphi_1=2\pi$, i.e.,
\begin{equation}
\bigtriangleup\tilde{Y}_{0i}^{\pm}(x')
=-(a_i\pm b_i)+m_i
\end{equation}
\begin{equation}
\bigtriangleup\tilde{Y}_{Ni}^{\pm}(x)
=\xi^{-N}\left[(a_i\pm b_i)-m_i\right] .
\end{equation}
Hence we get the corresponding geodetic intervals when the
points $x'$ and $x$ are not on the symmetry axis of the Klein
bottlehole,
\[\sigma_{Ni}^{\pm}=\zeta_N\left[(\bigtriangleup\tilde{Y}_{Ni}^{\pm}\xi^N
-T)\xi^{-N}-(\bigtriangleup\tilde{Y}_{0i}^{\pm}
-T')\right]\]
\[=\zeta_N\left\{\xi^{-N}\left[\pm b_i (1 \mp (-1)^i \cos\varphi_2)-T\right] \right. \]
\begin{equation}
\left. +\left[\pm b_i (1 \mp (-1)^i \cos\varphi_2)+T'\right] \right\} .
\end{equation}
What is of most interest is the case when the points $x'$ and $x$
are also slightly off the throat in the $Y$ direction. Then
we obtain
\[\lambda_{Ni}^{\pm}(x,x')=\]
\begin{equation}
\pm 2b_i(1\pm(-1)^i\cos\varphi_2)
+(Y-T)-(Y'-T')\xi^N .
\end{equation}
In order to uncover the quantum instabilities that can take
place in the accelerating Klein bottlehole, it is useful to
introduce the concept of $N$th-polarized hypersurface, $H_N$,
i.e., that hypersurface which is formed by those events that
join to themselves through closed null geodesics by traversing
the Klein bottlehole $N$ times [21]. Its interest arises from
the fact that quantum vacuum polarization diverges on such
hypersurfaces. Since Klein bottleholes are nothing but a
topological generalization of Misner space, one should expect
$N$th-polarized hypersurfaces to exist in the accelerating hole
space with the symmetry of a Klein bottle. In fact, upon
collapsing the points $x'$ and $x$ together in (4.12) it follows
that there will be polarized hypersurfaces at times fixed by the
condition $\sigma_{Ni}^{\pm}=0$, and hence $\lambda_{Ni}^{\pm}=0$, i.e.,
at times
\begin{equation}
T_{H_N i}^{\pm}=
\pm\frac{\xi^{-N}+1}{\xi^{-N}-1}b_i(1\mp(-1)^i\cos\varphi_2) .
\end{equation}

There will be four chronology (Cauchy) horizons, $H_{i}^{\pm}$,
which appear as the limit as $N\rightarrow\infty$ of the
times $T_{H_N i}^{\pm}$ in accelerating Klein bottlehole space.
They will respectively nest the corresponding polarized
hypersurfaces defined at the times given by (4.13). On the
symmetry axis where $\varphi_2=\pi$ and $\varphi_2=0$, all
polarized hypersurfaces $H_{Ni}$ occur at the same time only
at $T=0$; away from this symmetry axis, one meets the
polarizaed hypersurfaces one after another beginning with
arbitrarily large $N$ and ending at $N=1$, as $T$ increases
if we are in region $2\pi-\varphi_2^c >\varphi_2 >\varphi_2^c$,
or as $T$ decreases if $-\varphi_2^c <\varphi_2 <\varphi_2^c$.

Each of the four different chronology horizons
nests a set of polarized hypersurfaces. The nesting
of hypersurfaces $H_{N1}$ in the chronology horizon $H_1^+$,
occurring at time
\[T_{H_1^+}=+b_1(1+\cos\varphi_2) ,\]
guarantees that, an observer entering the region of CTC's will
pass first through the chronology horizon $H_1^+$, and then
successively through the $H_{N_1}$'s, at which hypersurfaces the
observer would experience the strong peaks of vacuum polarization.
The same behaviour would also be expected for the chronology
horizon $H_2^+$, which occurs at the different time
\[T_{H_2^+}=b_2(1-\cos\varphi_2). \]

In the case of the other two horizons occurring at times
\[T_{H_1^-}=-b_1(1-\cos\varphi_2) \]
and
\[T_{H_2^-}=-b_2(1+\cos\varphi_2), \]
the observer will first pass through the corresponding
successive polarized hypersurfaces ($H_{N_1}^{-}$ and
$H_{N_2}^{-}$, respectively) and then enters the given
chronology horizon.

Anyway, we see that the kind of semiclassical instabilities
which
were present in wormholes and ringholes are also present
in Klein bottleholes. We could however re-define the vacuum
corresponding to the accelerating holes in Euclidean space
to make it self-consistent [6]. With this new unique
vacuum, the renormalized energy-momentum tensor should
turn out to vanish everywhere, in principle so avoiding
any instabilities of the accelerating Klein bottlehole
or actually any of the time machines obtained by
topologically generalizing Misner space. Nevertheless,
according to Kay, Radzikowski and Wald [13], this will only
guarantee quantum stability on the regions just up to the
chronology horizon, since such horizons lose their
physical meaning also in the new vacuum, and the quantum
divergence problem would still remain.

In spite of the failure to keep quantum-mechanically stable
macroscopic time machines, one still could make it possible to
avoid quantum instabilities in the above accelerating holes if
we consider holes obtained as generalizations from a modified
Misner space where the period of the closed spatial direction
becomes time-dependent and given by $2\pi T$ [7]. Using then
automorphic fields [26] to compute the Hadamard function one
obtains a quantization condition for time $T=(N+\alpha)T_0$,
where $0\leq\alpha\leq\frac{1}{2}$ is the automorphic constant
and $T_0$ is a minimum constant time whose most sensible value
would probably situate on the Planck scale. Translating into the
languaje used above, it follows that the condition for the
existence of the $N$th-polarized hypersurfaces on which quantum
polarization of vacuum diverges should in this case imply
\[\frac{(1+\xi^N)b_i}{(1-\xi^N)(N+\alpha)}(1\mp(-1)^i\cos\varphi_2)
=T_0=Const. ,\]
which the system obviously cannot fullfil, unless $\xi$, $b_i$,
$\varphi_2$ and $N$ take on specific constant values.
Therefore, no $N$th-polarized hypersurfaces could exist in any
of the accelerating holes which are generalizations from this
modified Misner space. However, on such generalizations only
are possible CTC's with sizes of the order the Planck time [7]
and their chronology horizons will possess nonzero width,
also on the Planck scale.

\section{\bf Conclusion}

Using a convenient ansatz for the geometric parameters that
describe a Klein bottle, we have obtained exact solutions to the
associated Einstein equations for two-dimensional sections in
the vacuum case. These solutions possess apparent horizons at
fixed valued of the two angular variables used to describe the
Klein bottle. Starting with these solutions, we have constructed
a spacetime which represents a Klein bottlehole tunneling that
connects two asymptically flat large regions by shortcutting
spacetime, and found its embedding conditions in flat space. The
latter conditions require that, at or near the throat, the
embedding surface flares toward the two extreme values of the
radius of the transversal section of the Klein bottle tube, and
ultimately correspond to a rather complicated distribution of
ordinary and exotic matter around the hole throat. The matter
distribution allows, however, the existence of itineraries
through the tunnel along which an observer could avoid finding
regions with negative energy density, and gives rise to
different lensing effects of the Klein bottlehole's mouths. We
then constructed a time machine out of this spacetime hole and
obtained its metric by allowing one of the hole's mouths to move
relative to the other (see however [27]). This will allow that,
at sufficiently late times CTC's arise in some nonchronal region
by relativistic time dilation.

There are four different chronology horizons in the resulting
accelerating Klein bottlehole. Roughly speaking, the chronology
horizons can be regarded to be like light cones developed from
points of the original space. The four distinct horizons nest
four correspondingly different classes of polarized hypersurfaces
on which vacuum quantum fluctuations diverge, so making the time
machine quantum-mechanically unstable. This result is inscapably
obtained whenever our nonorientable spacetime construct can be
regarded as a topological direct generalization from usual Misner
space, where nearly the same kind of instability also occurs. However,
we have argued that if the nonorientable accelerating Klein
bottlehole, and actually any other such topological generalizations
(e.g. accelerating wormholes and ringholes) are instead taken to
be similar generalizations from the recently proposed modified
Misner space [7] (i.e. Misner space with a time-dependent period
of the closed spatial direction), then the calculation of the
regularized two-point Hadamard function implies a quantization of
time that ultimately prevents the existence of polarized
hypersurfaces, and hence leads to a quantum-mechanically stable
time machine. The price to be paid for this is to have to
renounce to the possibility of having time machines which
would produce CTC's involving large time displacements. In
fact, when time is quantized in very small steps $T_0$, the
resulting CTC's involve only time intervals of order $T_0$ [7].
It is in this sense that stabilization of time machines does
not induce any violation of semiclassical chronology
protection conjecture [8]: quantization of time is simply not
included in the conjecture.

Quantum spacetime foam can be thought to have a number of
components, such as wormholes, virtual black holes [28], etc,
among which quantum time machines inducing local violations of
causality and orientability (i.e. accelarating Klein bottleholes
arising as generalizations from modified Misner space) seem to
be most necessary if the foam is defined in terms of minimal
values of time and length at nearly the Planck scale [19].
Whether or not a future civilization will be able to extract,
grow up and mantein one such time machines out from the foam in
such a way that the minimum time $T_0$, and hence CTC's be also
scaled to large values is a question that only future
development might answer.

\acknowledgements

\noindent For useful comments, the authors thank
A. Ferrera and G.A. Mena Marug\'{a}n of IMAFF. This research was
supported by DGICYT under Research Projects No. PB94-0107 and
PB97-1218.

\vspace{1cm}

\begin{center}
{\bf Legend for Figure}
\end{center}

\noindent Fig. 1. Cartesian coordinates on the two-dimensional
Klein bottle. Any point $P$ on the Klein bottle surface can be
labeled by parameters $a,b,\varphi_1,\varphi_2$. If $P$
correspons to the interval $0\leq\varphi_1\leq2\pi$ (as
displayed on the figure), then the reference frame to fix the
given point is that with origin at $O_1$ (with parameters
$a_1,b_1,m_1,r_1$ in the main text), and if $P$ corresponds to
the $\varphi_1$-interval, $2\pi<\varphi_1 <3\pi$, associated to
the bottle region which makes it nonorientable, then it would be
given in terms of the reference frame with origin at $O_2$ (with
parameters $a_2,b_2,m_2,r_2$ in the main text).

\end{document}